\documentclass[12pt]{iopart}
\usepackage{graphicx}
\begin{document}

\def\Journal#1#2#3#4{{#1} {\bf #2}, #3 (#4)}

\def\NCA{\em Nuovo Cimento}
\def\NIM{\em Nucl. Instrum. Methods}
\def\NIMA{{\em Nucl. Instrum. Methods} A}
\def\NPB{{\em Nucl. Phys.} B}
\def\NPA{{\em Nucl. Phys.} A}
\def\PLB{{\em Phys. Lett.}  B}
\def\PRL{\em Phys. Rev. Lett.}
\def\PRC{{\em Phys. Rev.} C}
\def\ZPC{{\em Z. Phys.} C}
\def\JPG{{\em J. Phys.} G}
\def\EPJ{{\em Eur. Phys. J.} C}

\title{Resonances at RHIC}

\author{Zhangbu Xu\dag
\footnote[3]{mailto: xzb@bnl.gov}
}

\address{\dag\ Physics Department, Brookhaven National Laboratory,
Upton, NY 11973}

\begin{abstract}
In this report, we discuss the measurement of the hadronic decay
modes of resonances in relativistic heavy-ion collisions,
emphasizing on RHIC results. The study of resonances can provide:
(1) the yield and spectra of more particles with different  
properties to study whether the system is in thermal equilibrium;
(2) an independent probe of the time evolution of the source from
chemical to kinetic freeze-out and detailed information on
hadronic interaction at later stage; (3) the study of 
medium effects at a late stage of the collisions; (4) a measurement
of flavor and baryon/meson dependence of particle production at
intermediate $p_{T}$ ($>2$ GeV/$c$).
\end{abstract}




\subsection{Hadron Production in Heavy-Ion Collisions}

Historically, the discovery of several important resonances
($\Sigma(1385)$, $K^{\star}(892)$, $\rho^{0}(770)$, $\eta$, etc.)
helped confirm the quark model (the Eightfold way) in the early 60's
\cite{alvarez}. Among more than four thousand particles discovered,
most of them are resonances \cite{rafelski3}. In a simple system such
as $e^{+}e^{-}$ collisions at $\sqrt{s}\!=\!91$ GeV, the production
characteristics of 46 different particles have been measured
\cite{pdg}.  With same way of counting, only 13 particles ($\pi^{\pm},
\pi^{0}, K^{\pm}, K^{0}, \rho^{0}, \eta, \phi, K^{\star0}, J/\Psi, p,
\Lambda, \Xi, \Omega$) have been measured in ultrarelativistic
heavy-ion collisions up to date.

The soft processes of QCD hadronization and interaction are not
calculable with a perturbative approach and instead rely on
phenomenological models to describe the particle production. In a
simple system, such as $e^{+}e^{-}$ collisions, the string and cluster
fragmentation can be used in Monte Carlo Models or in an empirical
model to describe the data \cite{jypei}. In heavy-ion collisions,
sufficient interactions among particles after their creation may lead
to thermalization and therefore a thermal model could be used to
describe the data. In fact, thermal models have been able to
successfully fit the data and extract thermal parameters such as
temperature and chemical potential at AGS, SPS and RHIC energies
\cite{thermalmodel}. In these models, a large fraction of the measured
yields of stable particles comes from the feeddown of resonance
decays. In some cases, it is the dominant contribution to the measured
value. For example, in JETSET and thermal models, less than 20\% of
measured pions (30\% kaons) do not originate from
resonances~\cite{jypei,thermalmodel}, while the resonance production
in elementary collisions is dominated by primary production. It is
therefore important to measure resonances to constrain the feed-down
and to continue testing the validity of thermal models. The same
argument is applicable to transport models (such as UrQMD
\cite{bleicher}, AMPT \cite{ampt}) and hydrodynamic models.

\begin{figure}[ht]
\begin{center}
\includegraphics*[keepaspectratio,scale=0.35]{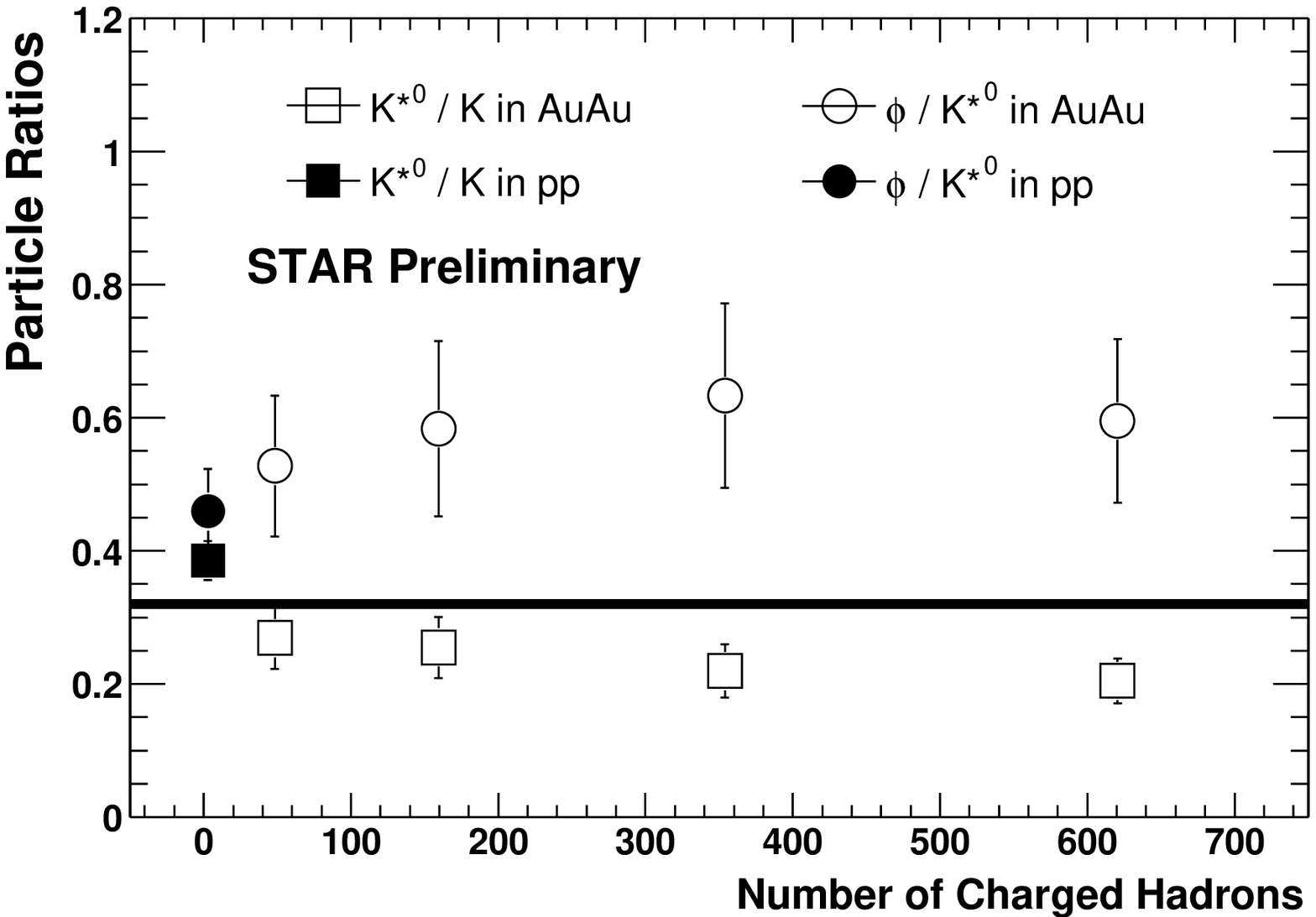}
\includegraphics*[keepaspectratio,scale=0.28]{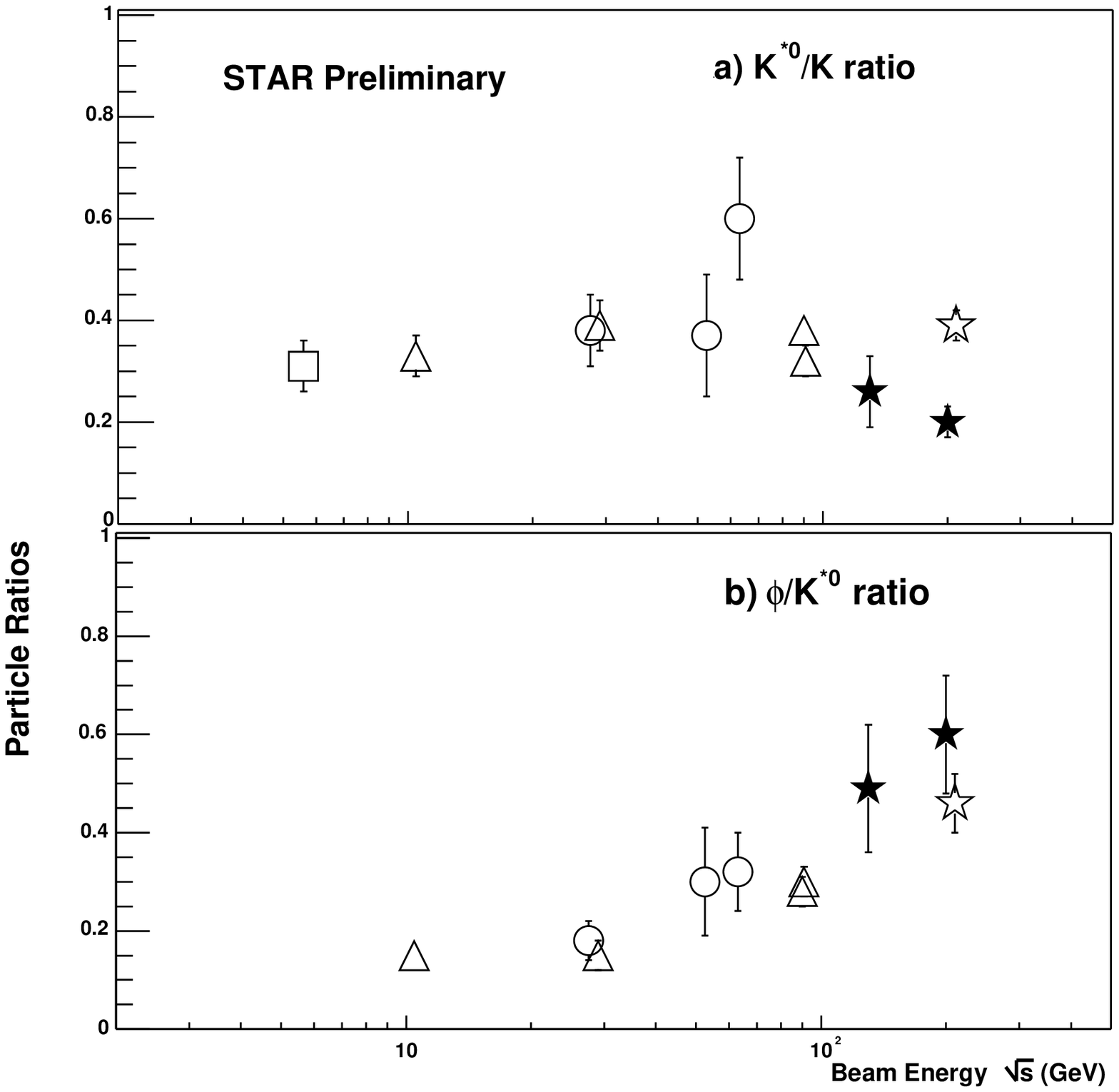}
\caption{Left panel: $K^{\star0}/K$ ratio and $\phi/K^{\star}$
ratio as function of charged hadron multiplicity in Au+Au
collisions at $\sqrt{s_{_{NN}}}\!=\!200$ GeV. The filled symbols
are results from pp collisions. The line is a prediction from
thermal model \cite{thermalmodel}. Right panel: $K^{\star0}/K$
ratio and $\phi/K^{\star}$ ratio as function of beam energy in
$e^{+}e^{-}, pp, \bar{p}p$ and Au+Au collisions. Filled symbols
are for Au+Au collisions and the open symbols are for elementary
collisions.} \label{Fig:ratios}
\end{center}
\end{figure}

The first $K^{\star}$ measurement by STAR on $K^{\star}(892)/h^{-}$
\cite{kstar} at $\sqrt{s_{_{NN}}}\!=\!130$ GeV/c is consistent with
thermal fit \cite{thermalmodel} with the combined statistical error
and systematic error of the order of 30\%.  The ratio $K^{\star0}/K$
is less model dependent than $K^{\star0}/h^-$ since both particles
have similar quark content and differ only in their spin and mass. The
right panel in Fig.~\ref{Fig:ratios} shows that this ratio is
independent of beam energy in $pp$, $\bar{p}p$ and
$e^{+}e^{-}$. Thermal model predicts this ratio to be 0.32 at chemical
freeze-out at RHIC \cite{thermalmodel}. The STAR results on
$K^{\star}(892)/K\!=\!0.26\!\pm\!0.03\!\pm\!0.07$ \cite{haibin} in
central Au+Au collisions at $\sqrt{s_{_{NN}}}\!=\!130$ GeV/c and
$K^{\star}(892)/K\!=\!0.20\!\pm\!0.01\!\pm\!0.03$ \cite{haibin} in
central Au+Au collisions at $\sqrt{s_{_{NN}}}\!=\!200$ GeV/c with
improved statistical and systematic uncertainty \cite{kstar} show that
this ratio does have a centrality dependence and is significantly
lower than thermal model prediction or results from $pp$ collisions
measured by the same experiment. Siginificant decrease of
$\Lambda(1520)$ in Au+Au collisions with respect to $pp$ and thermal
calculation are also observed \cite{ludovic}.

Whether the remarkable success of the statistical model may simply
mean phase-space dominance and the ``temperature'' and ``chemical''
potentials are nothing but Lagrange multiplier characterizing the
phase-space integral \cite{koch} is still unclear. The existence of
flow is considered a strong evidence that certain amount of
rescattering takes place among the particles in the bulk matter formed
in heavy-ion collisions. However, as stated in Ref.  \cite{koch}, ``to
which extent they are sufficient to form matter in the Boltzmann sense
is, however, not clear''. The strong rescattering and regeneration of
resonances after chemical freeze-out as discussed are direct evidence
of rescattering of particles in the bulk matter formed in heavy-ion
collisions. It is difficult to imagine that such interaction will not
be stronger at even earlier stage.

\subsection{Time Evolution of the System}
Resonances which decay into strongly interacting hadrons inside the
dense matter are less likely to be reconstructed due to the
rescattering of the daughter particles. On the other hand,
resonances with higher $p_T$ have a larger probability of decaying
outside the system and, therefore, are more likely to be
reconstructed. Examples of how these measurements can be used as a
signature of freeze-out dynamics are shown in Refs.
\cite{kstar,rafelski,broniowski}.

Recent transport model (UrQMD) calculation shows significant
modification of resonance population and $p_{T}$ spectrum at
mid-rapidity due to rescatterings \cite{bleicher}. In addition, the
resonance yield could be increased during the rescattering phase
between chemical freeze-out (vanishing inelastic collisions for both
stable particles and resonances) and kinetic freeze-out (vanishing
elastic collisions) via, for example, the elastic process
$\pi{K}\!\rightarrow\!{K^{\star0}}\!\rightarrow\!\pi{K}$. This
regeneration mechanism partially compensates for resonance decays
during a possible long expansion of the system and increases the
observable ratio $K^{\star0}/K$. The $\pi\pi$ scattering is the
dominant process in destroying the reconstruction of $K^{\star}$ at
later stage while the $K\pi$ reaction is the dominant channel of
$K^{\star}$ regeneration. The $\pi\pi$ cross section is larger than
the $K\pi$ cross section by a factor of $\sim$5 \cite{urqmd} around
the resonances.  This means that at sufficiently late stage, the
rescattering of the daughters via the scattering of decay pions will
be much larger than the regeneration process via the $K\pi$
reaction. Therefore, significant effect from these rescattering should
be detectable experimentally.

It has been shown in this conference that there is a significant
decrease of $K^{\star}/K$ \cite{haibin} and
$\Lambda^{\star}/\Lambda$ \cite{ludovic} ratios while $\phi/K$
ratio is independent of beam specie and centrality
\cite{ma,phi}. This indicates that the yields are related to the
cross section of the interactions between their
decay daughters and the surrounding hadrons. It can also be seen
from the $p_T$ spectra of $K^{\star}$ and $\phi$ as shown in right
panel of Fig.~\ref{Fig:spectraheavy} that the low $p_T$ part of
$K^{\star}$ is suppressed relative to $\phi$ and other particles.
Resonances being discussed are $\phi(1020)$,
$K^{\star\pm}(892)$, $\Lambda$(1520), $\Sigma$(1385),
$\Delta^{++}$(1232), $\rho^{0}(770)$ and $f_0(980)$. The ordering
of the lifetime is
$\phi(1020)\!>\!\Lambda(1520)\!>\!\Sigma(1385)\!>\!K^{\star}(892)
\!>\!f_0(980)\!>\!\Delta^{++}(1232)\!>\!\rho^0(770)$.
The ordering of their regeneration cross section is
$\Delta^{++}(1232)\geq\rho^0(770)\!>\!f_0(980)\!>\!K^{\star}(892)
\!>\!\Sigma(1385)\!>\!\Lambda(1520)\!>\!\phi(1020)$.
Meanwhile, the rescattering of daughters in all cases is dominated
by either $\pi\pi$ or $p\pi$. The ordering of their mass
difference to ground state is $\Sigma(1385)/\Lambda \!<\!
\Delta^{++}(1232)/p \!<\! K^{\star}(892)/K \!<\!
\Lambda(1520)/\Lambda \!<\! \rho^0/\pi \!<\! f_0/\pi$. If the
rescattering is the dominant process, the resonance yields and
$p_{T}$ spectra will be related to the lifetime and cross
sections. On the other hand, if the dominant processes are both
the  regeneration and rescattering with a long time expansion
which tend to equilibrate the yields at kinetic freeze-out, the
resonance yields and $p_{T}$ spectra will be related to the
Boltzmann factor. By systematically comparing the yields and $p_T$
distributions of resonances with other particles it may be
possible to distinguish different freeze-out conditions; {\it
e.g.} sudden freeze-out \cite{rafelski,broniowski} or a slow
hadronic expansion \cite{bravina}. From the preliminary results of
$K^{\star}/K^-, \Lambda(1520)/\Lambda, \rho^0/\pi^-, f_0/\pi^-$
and $\phi/K^-$ \cite{pfachini,haibin,ludovic,ma} measured by STAR
and the comparison with thermal model predictions at chemical
freeze-out \cite{thermalmodel}, it seems that the ratios follow
the regeneration cross section. This implies that large
rescattering of daughters happens after chemical freeze-out.
However, the regeneration is not enough to compensate the loss to
equilibrate the yield due to fast freeze-out and/or decreasing
temperature.

It has been noted that \cite{kstar,rapp,shuryak} due to the
overpopulation of $\pi$'s at kinetic freeze-out, the resonances
having a $\pi$ daughter will have higher yield due to additional
pion chemical potential. This will change the ordering of the
particle ratios mentioned above. With these precise measurements,
it is now clear that the yield of resonances are not consistent
with thermal predictions \cite{thermalmodel,broniowski} at
chemical freeze-out. The interpretations of these results are
being actively pursued as seen in these proceedings and recent
publications. The results from the combination of a thermal model
as the initial condition together with the UrQMD transport model
compared to the measured yields and spectra of resonances and
other particles will constrain both the thermal model and
evolution of the system \cite{bleicher,giorgio}. Additional
results from $\Delta^{++}$ and $\Sigma(1385)$ are coming soon from
STAR as well \cite{ansevil}. STAR is also carrying out the studies
of resonance elliptic flow ($v2$) and decay angular distribution
($\cos{\theta^{\star}}$) which are sensitive to the geometry of
the system and the rescattering \cite{pfachini,haibin}. These
studies, however, require much more statistics and better
systematical error than the yields and $p_{T}$ spectra.

\subsection{Medium Effect in Hot and Dense Matter}

\begin{figure}[ht]
\begin{center}
\includegraphics*[keepaspectratio,scale=0.35]{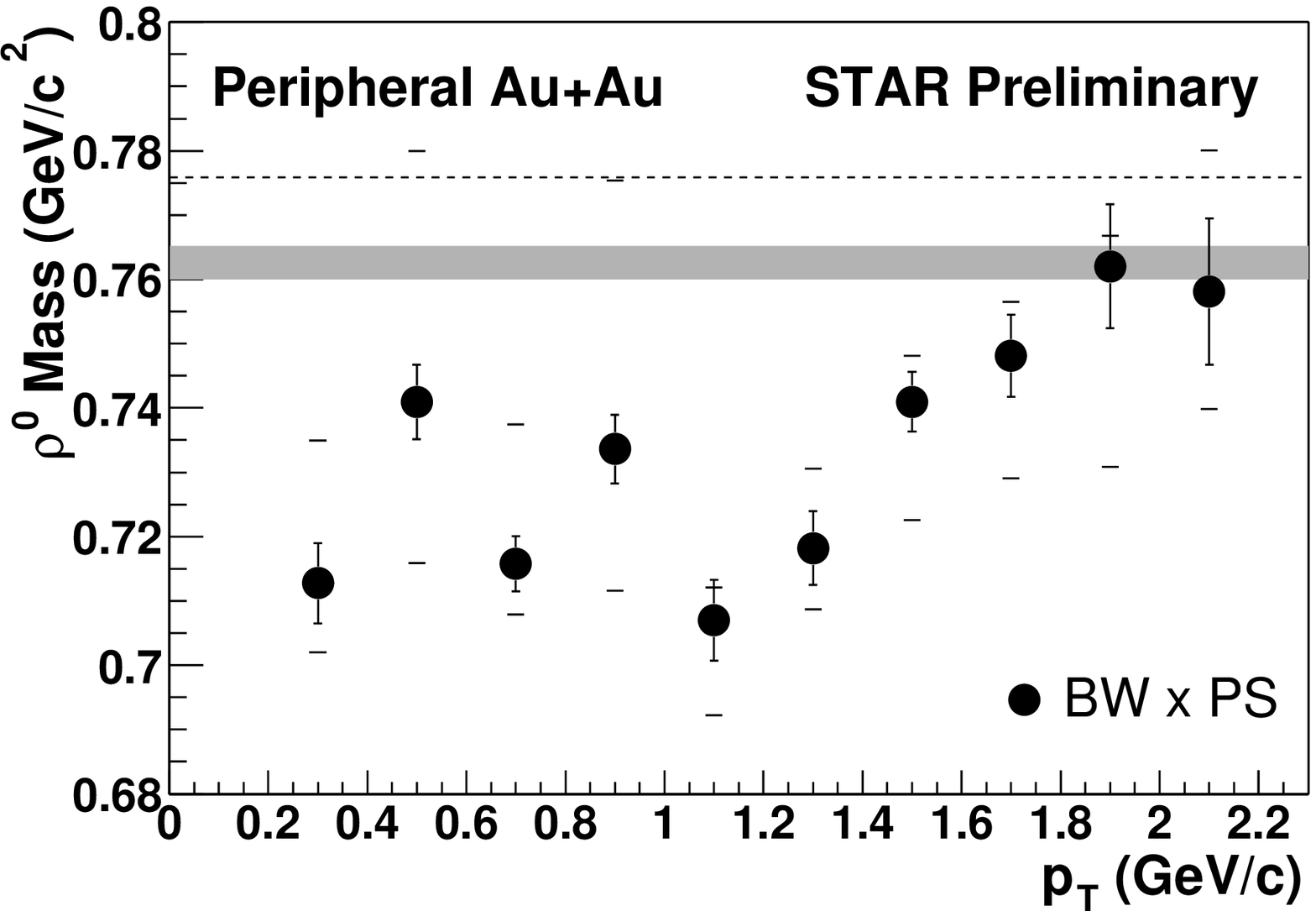}
\includegraphics*[keepaspectratio,scale=0.35]{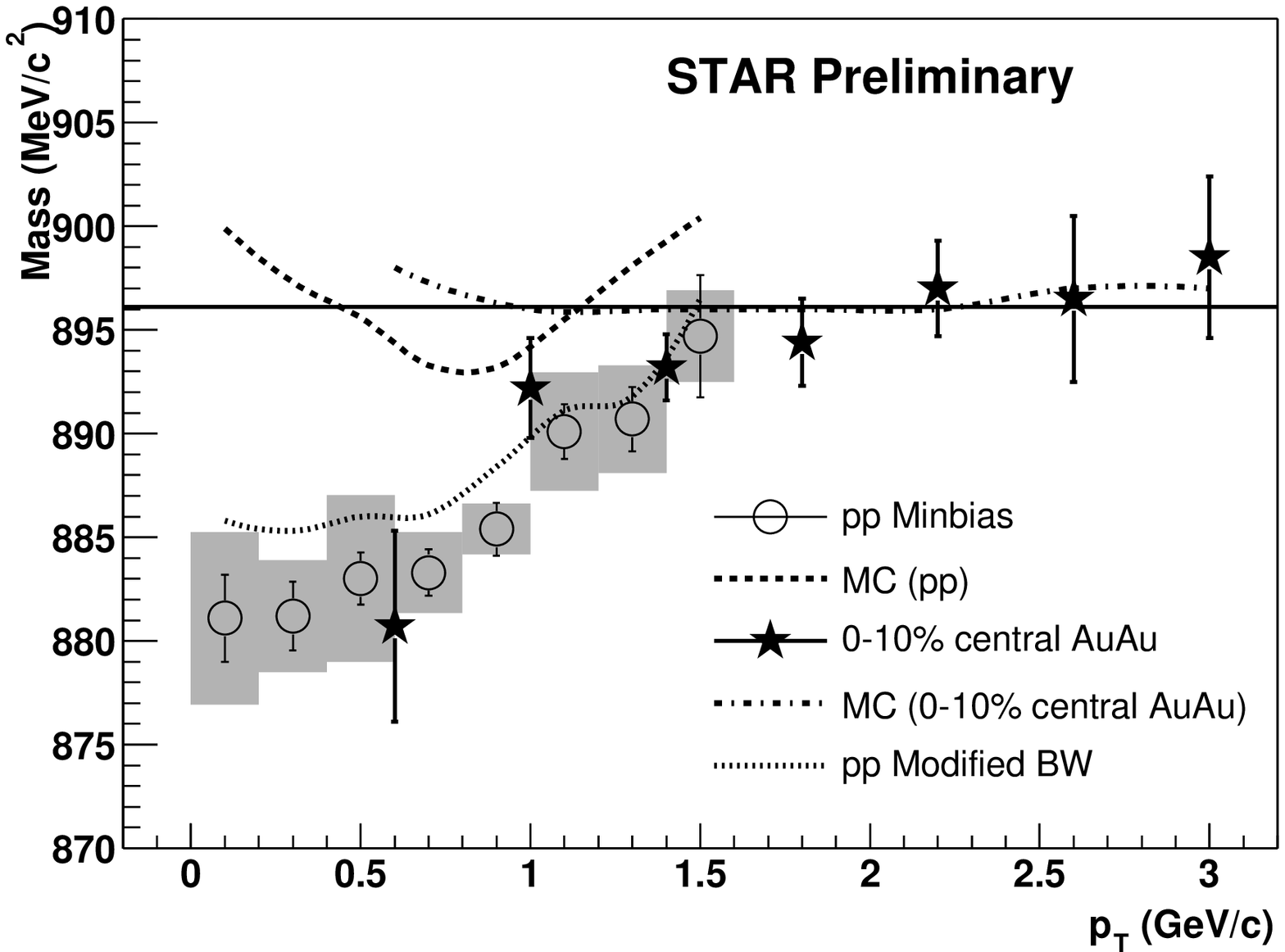}
\caption{Left panel: $\rho^{0}$ mass as function of $p_T$ in
Au+Au collisions with centrality of 40-80\%. The filled circles
are results from the fit to a Breit-Wigner function modified by a
phase-space factor. Right panel: $K^{\star0}$ mass vs $p_T$ in pp
and Au+Au collisions.} \label{Fig:mass}
\end{center}
\end{figure}

Among the proposed signals of phase transitions in hot, dense nuclear
matter produced via relativistic heavy-ion collisions are
modifications of the meson resonance production rates and their
in-medium properties \cite{rapp}.  When the resonance lifetime is
comparable to the evolution time scales of the phase transition, the
measured properties associated with the resonance (such as mass,
width, branching ratio, yield and transverse momentum ($p_T$) spectra)
will depend upon collision dynamics and chiral properties of the
medium at the high temperature and high energy density. The famous
result is the enhancement of dileptons in the medium mass range around
500 MeV by CERES/SPS \cite{ceres}. Models with in-medium mass
modification of $\rho^{0}$ were proposed to explain the large
enhancement. The resonances measured in the hadronic decay channel
probe only the late stages of the collision. It has been argued that
significant effect on the reconstructed resonance mass can exist in
such late stage due to phase space \cite{shuryak,kolb}, interference
\cite{longacre}, rescattering \cite{bleicher} and dynamical effects
\cite{shuryak,rapp}. As shown in \cite{pfachini,haibin} and here in
Fig.~\ref{Fig:mass}, STAR has observed significant mass shifts in both
$\rho^{0}\!\rightarrow\!\pi^{+}\pi^{-}$ and
$K^{\star}\!\rightarrow\!\pi K$ decays \cite{haibin,pfachini}. The
mass positions are consistently lower than the PDG (Particle Data
Group \cite{pdg}) values and $p_T$ dependent with a large discrepancy
at lower $p_T$ and with the measured values approaching the PDG values
at higher pT. It was found that the $\rho^{0}$ mass at low $p_T$ was
about -30 MeV/$c^{2}$ and -70 MeV/$c^{2}$ below the PDG values in $pp$
and peripheral Au+Au collisions while the $K^{\star0}$ value is about
-15 MeV/$c^{2}$ below the PDG values. The statistical error of
$K^{\star0}$ in Au+Au collisions is too large to be conclusive. The
mass and width of other resonances ($\phi,\Sigma^{\star}$ and
$\Lambda^{\star}$) and strange particles
($K_{S}^{0},\Lambda$) are consistent with the PDG values within the
STAR experimental resolution \cite{ma,ludovic,hui}.

The first question one should ask is whether the mass shift of
these particles (especially the copiously produced $\rho^{0}$
vector meson) has been observed anywhere. In fact, previous
measurements of the $\rho^0$ meson in hadronic $Z^0$ decays by the
OPAL, DELPHI and ALEPH collaborations indicate that the $\rho^0$
line shape is considerably distorted from a relativistic p-wave
Breit-Wigner shape, especially at relatively low momentum in
multipion systems \cite{opal,delphi,aleph,lafferty}. A mass shift
that may be -30 MeV/$c^{2}$ or larger was observed
\cite{opal,aleph}. The OPAL experiment at LEP has reported
$\rho^\pm$ mass shifts of -10 to -30 MeV/$c^2$ in the position of
the maximum of the resonance consistent with the observed $\rho^0$
mass shift \cite{opal}. In the paper by the DELPHI collaboration,
the low momentum region, below 2.3 GeV/$c$ was excluded and a
Breit-Wigner shape was used to fit the $\rho^{0}$
\cite{delphi,lafferty}. Although the fitted mass was some five
standard deviations below the PDG value, rates for the $\rho^{0}$
were nonetheless extracted using a conventional Breit-Wigner shape
\cite{delphi,lafferty}. The DELPHI collaboration reported the
$\rho^{0}$ mass as $757 \!\pm\! 2$ MeV/$c^2$ \cite{delphi}, which
should be compared to the PDG average value of $775.9 \!\pm\! 0.5$
MeV/$c^{2}$ from $\tau$ decays and $e^{+}e^{-}$ interactions. The
ALEPH collaboration also reported that the $\rho^{0}$ resonance in
the data seemed to be shifted towards lower masses \cite{aleph}.

In pp collisions, the $\rho^0$ meson has been measured at
$\sqrt{s}\!=\!$
 27.5 GeV \cite{na27}. This is the only pp
measurement used in the hadroproduced $\rho^0$ mass average
reported in the PDG. In this measurement, a $\rho^0$ mass of
0.7626 $\!\pm\!$ 0.0026 GeV/$c^2$ was obtained from a fit by a
relativistic p-wave Breit-Wigner function times the phase space
\cite{na27}. However, if the $\rho^0$ is fit only to a
relativistic p-wave Breit-Wigner function, a mass shift of -30
MeV/$c^2$ is observed.

In what follows, I would like to discuss some of the predicted
effects on the mass and width of the $\rho^{0}$ in the medium.

\subsubsection{Phase Space and Rescattering}

If the resonance observed is generated from scattering (e.g.
$\pi^{+}\pi^{-}\!\rightarrow\!\rho$, $K\pi\!\rightarrow\!
K^{\star}$), then the mass distribution will have contribution
from the phase space population of the parents (which are also
daughters). This distribution is characterized by the Boltzmann
distribution in a thermal system that tends to populate the low
invariant mass of the resonances and therefore effectively shifts
the resonance peak to lower values. Since in the invariant mass
distribution (without contamination from particle
misidentification often present in the experiments) the parents
(daughters) distribution represents the phase space and the
experimental acceptance and efficiency, the $\rho^{0}$ mass is
often obtained by fitting the same event distribution of
$\pi^{+}\pi^{-}$ to
\begin{equation}
BG + PS \times BW \!=\! BG + BG \times BW \!=\! BG(1 + BW),
\end{equation}
where BG corresponds to the background, PS is the phase space
factor and BW is the relativistic p-wave Breit-Wigner (e.g. Ref.
\cite{opal,delphi,na27} which was referred to in PDG). However,
this has a strong assumption that all resonances are created by
final state rescattering. In elementary collisions ($e^{+}e^{-},
pp$ and $\bar{p}p$), some of the resonances are directly from
string fragmentation which should not be modified by the phase
space population of other particles. In heavy-ion collisions, such
a fit is better justified since the system is much larger and
lives longer. This thermal factor has been discussed in several
papers \cite{shuryak,kolb}. Usually, this constribution to the
total mass shift is quite small ($\!<\!20$ MeV) and scales roughly
with $\Gamma^{2}$.

A more detailed study can be done using transport models, such as
UrQMD \cite{bleicher}. The left panel in
Fig.~\ref{Fig:kstarflavor} shows the $\rho^{0}$ mass for different
centrality from UrQMD in Au+Au collisions. The mass shift is
about -30 MeV/$c^2$ for central Au+Au collisions and vanishes for
peripheral collisions.

\subsubsection{Interference and Bose-Einstein Effect}

There are several sources of $\pi^{+}\pi^{-}$ correlations around
the $\rho^{0}$ mass:
\begin{itemize}
\item Direct $\rho^{0}$ production and its decay via $\rho^{0}\!\rightarrow\!\pi^{+}\pi^{-}$;
\item $\pi\pi$ scattering via $\rho^{0}$ resonant state;
$\pi^{+}\pi^{-}\!\rightarrow\!\rho\!\rightarrow\!\pi^{+}\pi^{-}$;
\item Direct $\pi\pi$ scattering via $\pi^{+}\pi^{-}\!\rightarrow\!\pi^{+}\pi^{-}$;
\item Interference between previous two scatterings;
\item Other $\pi\pi$ waves (S-, P-, D-waves).
\end{itemize}
The interference has been studied in great detail in the coherent
$\rho^{0}$ photoproduction. The STAR Collaboration used the same
detectors and the same beam to trigger on the $\rho^{0}$
photoproduced in ultraperipheral Au+Au collisions. The mass and
width are consistent with the PDG values and the interference
strength is consistent with other experiments \cite{upc}. It has
been argued that a similar interference may happen in $\rho^{0}$
production in $\pi^{+}\pi^{-}$ scattering \cite{longacre}. 
This, however, may not be compatible with vector moson dominance 
(VMD)~\cite{harada}. The
$\rho^{0}$ line shape in multihadronic $Z^0$ decay is addressed in
details in Ref. \cite{lafferty}. The final state Bose-Einstein
effect was introduced to explain the $\rho^{0}$ line shape
measured at LEP \cite{opal,delphi,aleph,lafferty,rapp}.

\subsubsection{$\rho^{0}$ Mass In-Medium Modification}

Although the effects mentioned above are interesting effects on
their own, in heavy-ion collisions the most interesting effect is
to identify the modification of particle (vector meson) properties
in the medium. The modification of the hadron spectral distribution
even at dilute systems are of particular interest since it will be
a precursor toward the chiral phase transition \cite{rapp}. A
couple of calculations using Brown-Rho scaling or dynamic width
broadening can qualitatively explain the mass shift
\cite{rapp,shuryak}. The preliminary results from STAR clearly
should be compared with theoretical calculations more
quantitatively in order to understand to what extent the
measurements reflect a true in-medium modification \cite{rapp}.
The hadronic channel measured by STAR coupled with future
measurements of the dilepton channel of vector mesons  will
provide an unique tool to study the in-medium effects as well as
the properties of the hot and dense matter.

\subsection{Mass or Baryon/Meson Dependence at Intermediate $p_T$}

\begin{figure}[ht]
\begin{center}
\includegraphics*[keepaspectratio,scale=0.30]{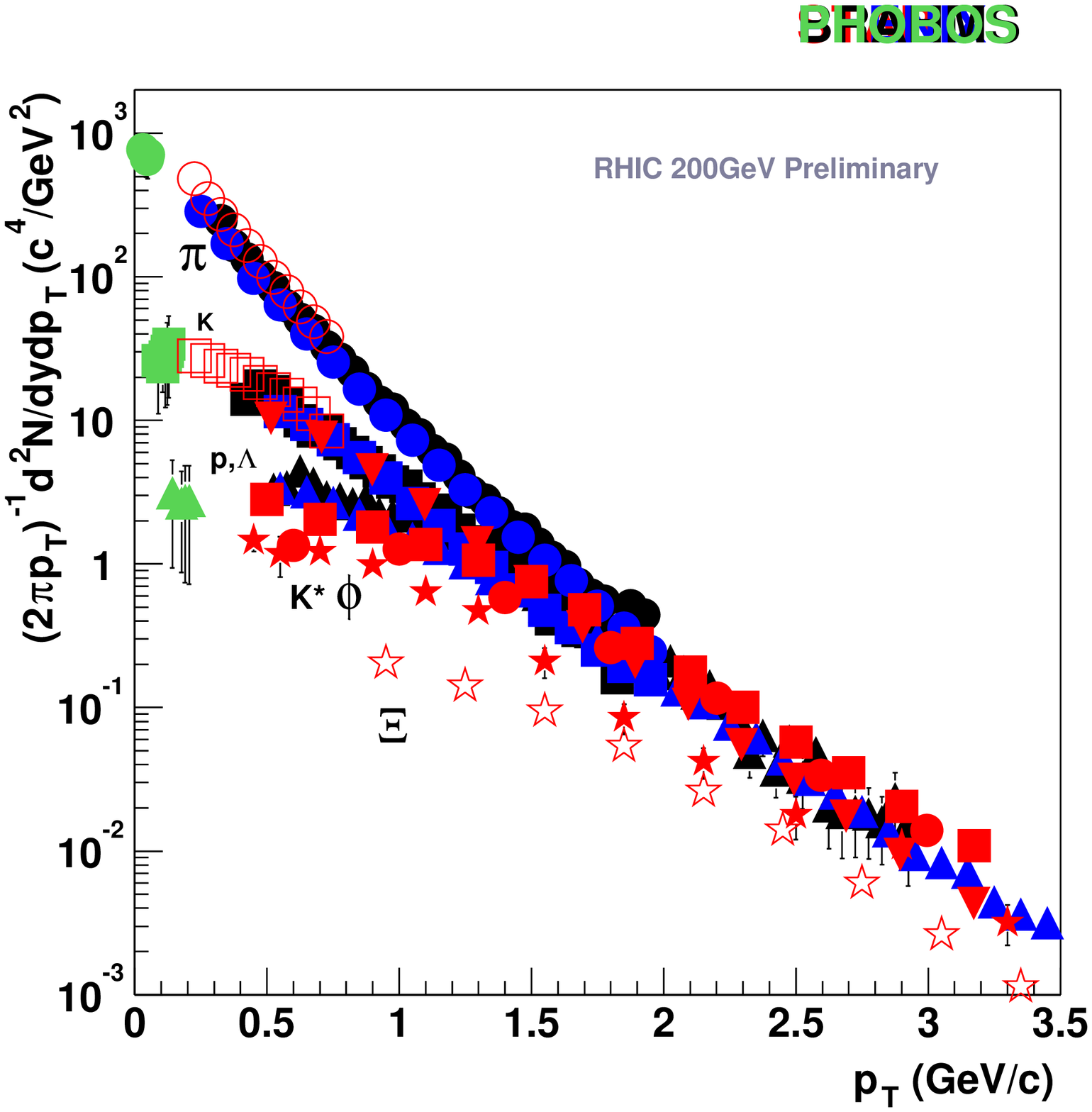}
\includegraphics*[keepaspectratio,scale=0.30]{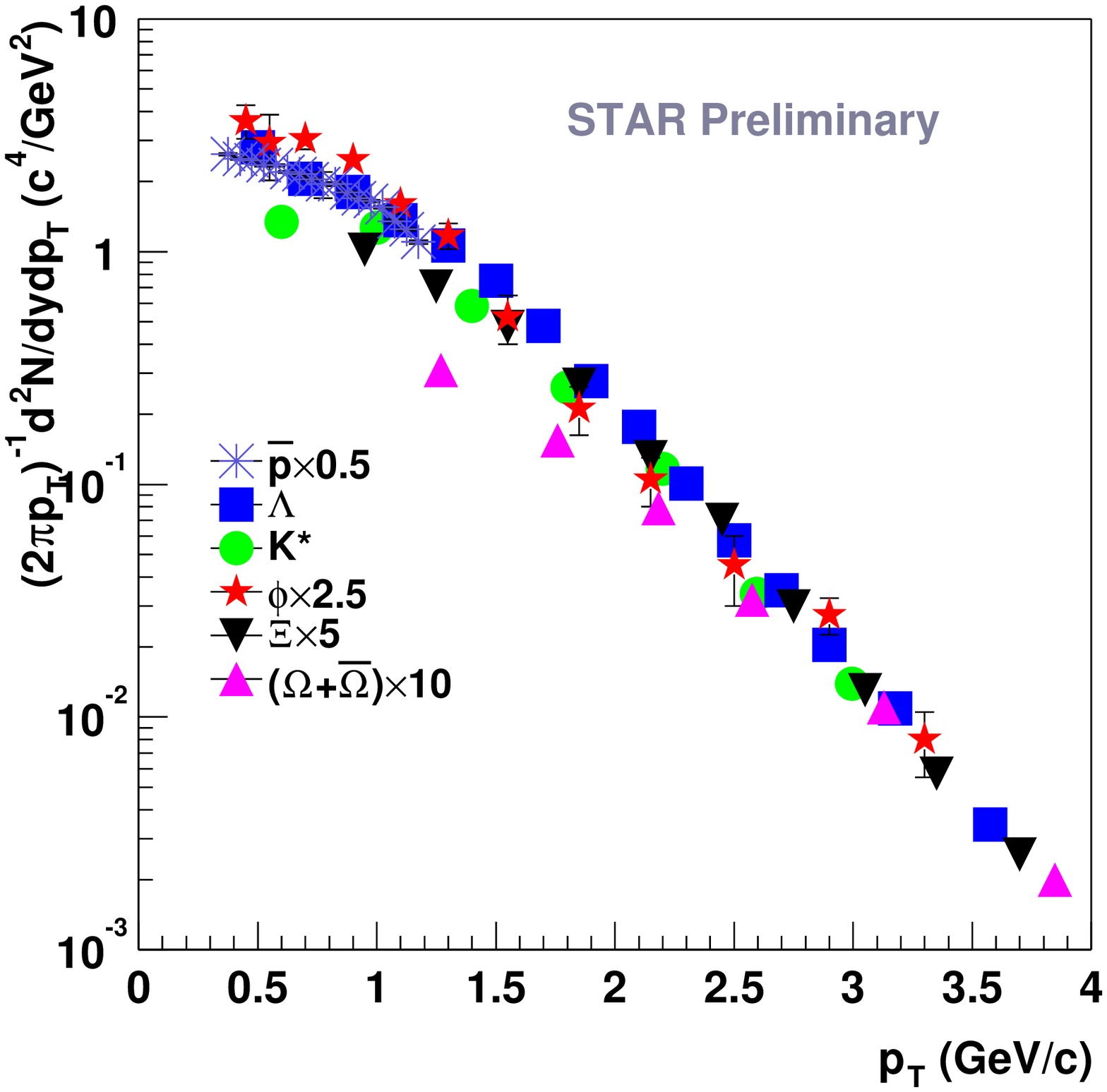}
\caption{$\pi, K, p, \Lambda, K^{\star0}, \phi$ and $\Xi$ $p_T$
spectra in central Au+Au collisions measured at RHIC. See text for details.}
\label{Fig:spectraheavy}
\end{center}
\end{figure}

A very prominent effect seen at RHIC is that the $p_T$ spectra of
different particles are very different from that of pp collisions at
intermediate $p_T$ \cite{vitev}. The effect has been attributed to
either a baryon-meson effect or a mass dependence from radial
flow. Unfortunately, the baryons are often heavier than mesons
(protons are heavier than pions and $\Lambda$'s are heavier than
kaons). The meson resonances such as $\phi$ and $K^{\star}$ are as
heavy as protons, but have the properties of mesons (two valence
quarks). Therefore, it is important to study the $p_T$ spectra of
these particles to distinguish baryon-meson or mass effect. The left
panel in Fig.~\ref{Fig:spectraheavy} depicts the particle spectra of
$\pi, K, p, \Lambda, K^{\star0}, \phi$ and $\Xi$ measured by all four
RHIC experiments \cite{ullrich,pfachini,haibin,ma,javier,hui}. The
right panel in Fig.~\ref{Fig:spectraheavy} shows the $p_{T}$ spectra
of particles with similar mass in central Au+Au collisions measured by
STAR.  The $\bar{p}$ spectrum includes the hyperon decay
feed-down. The spectra of $\bar{p}, \phi$ and $\Xi$ are scaled by a
factor of 0.5, 2.5 and 5 respectively. It is surprising to see that
these spectra are very close to each other in the intermediate $p_{T}$
range between 1 GeV/$c$ and 4 GeV/$c$. Although the rescattering at
late stages suppresses the $K^{\star}$ yield, the intermediate $p_T$
region ($p_{T}{}^{\!>\!}_{\!\sim\!} 1.5$ GeV/$c$) has a smaller effect
\cite{bleicher}. For $p_{T}\!<\!1.0$ GeV/$c$, we can see that
particles with different properties deviate from each other. For
example, $\phi$ has a smaller hadron cross section and therefore has
steeper slope at low $p_{T}$, while the $K^{\star0}$ daughters
rescatter in the hadronic matter and the low $p_{T}$ $K^{\star0}$'s are
suppressed. A study of the $\phi/K^{\star}$ ratio at intermediate
$p_T$ will reveal the production mechanism, e.g.  fragmentation and
recombination at partonic level \cite{mueller}.  From
Fig.~\ref{Fig:spectraheavy}, we see that the $\phi/K^{\star0}$ is
about 0.4. This is similar to the value in $pp$ collisions, as shown
in Fig.~\ref{Fig:ratios}. This indicates that the changes from $pp$ to
central Au+Au collisions are similar for the $\phi$ and
$K^{\star0}$. Further comparison between experimental data and
theoretical predictions on these spectra are needed.

\begin{figure}[ht]
\begin{center}
\includegraphics*[keepaspectratio,scale=0.35]{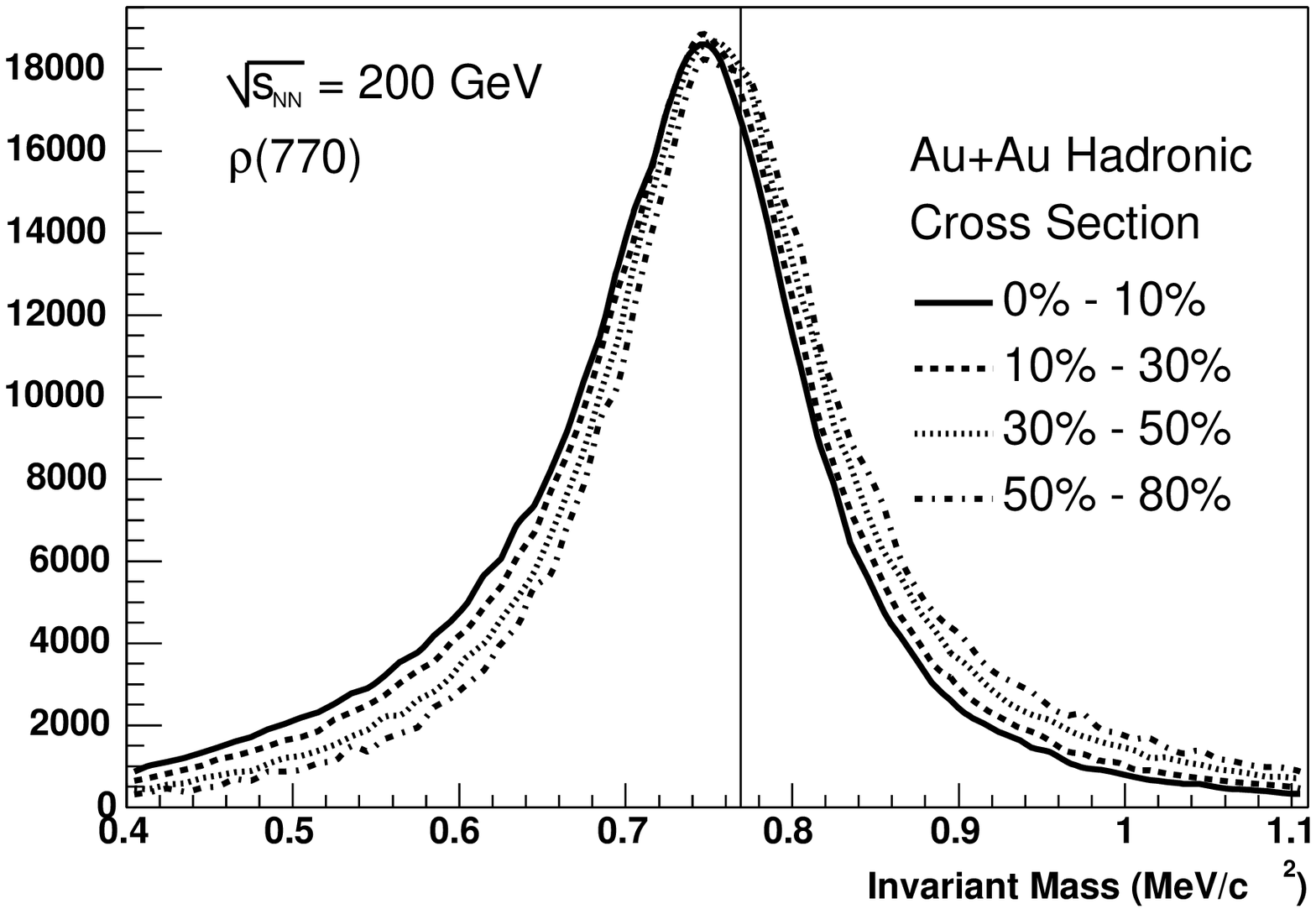}
\includegraphics*[keepaspectratio,scale=0.28]{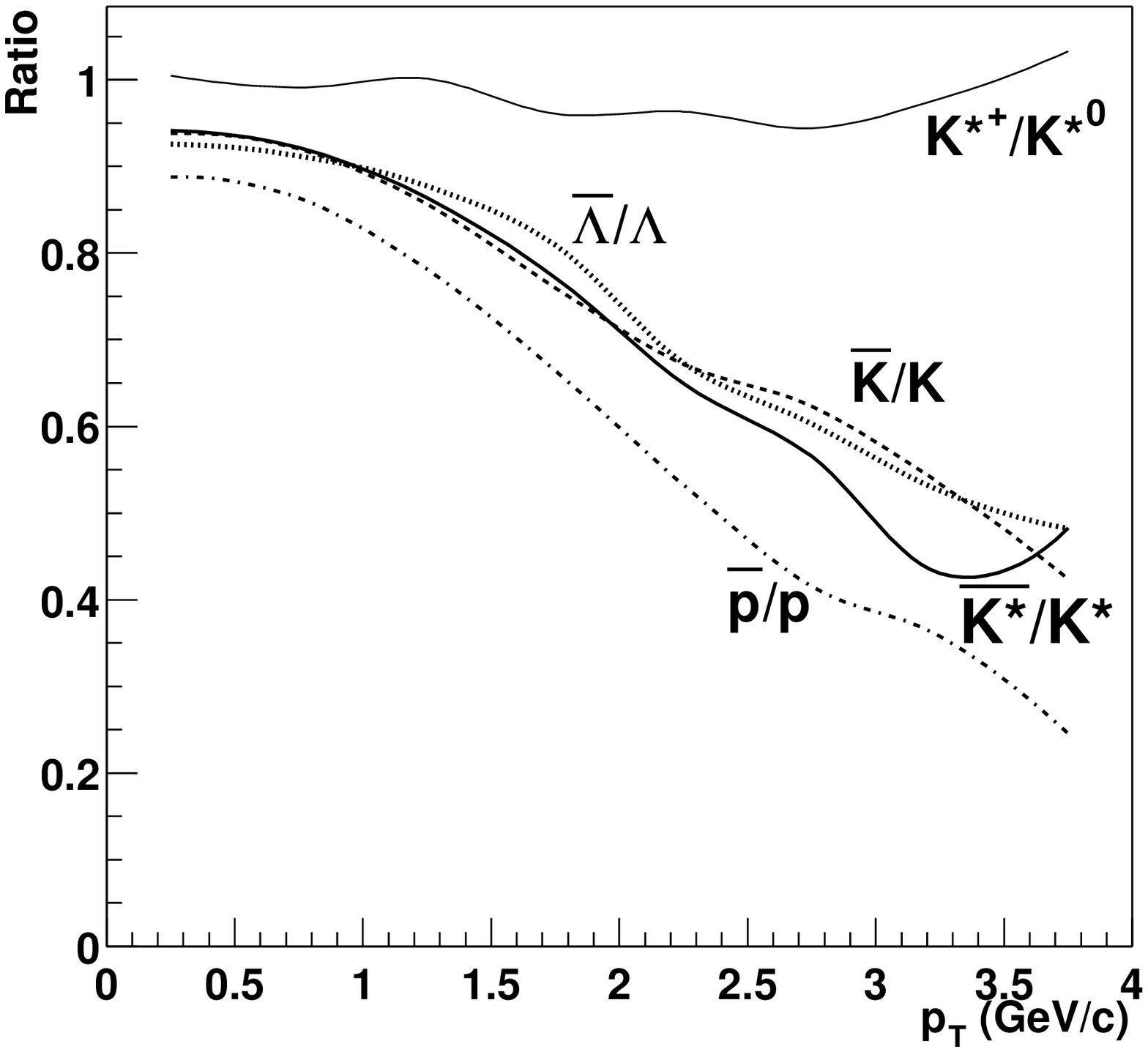}
\caption{Left: $\rho^{0}$ mass integrated over all $p_{T}$ from UrQMD 
  vs centrality in Au+Au collisions. Right: Effect of valence quark in
  particle ratio.  HIJING simulation of antiparticle to particle ratio
  as function of $p_{T}$. In case of $K^{\star}$, we have
  $\overline{K^{\star}}/K^{\star}$ (flavor) and
  $K^{\star+}/K^{\star0}$ (isospin).} \label{Fig:kstarflavor}
\end{center}
\end{figure}

Since there are more quark than antiquark jets and final hadrons
from gluon jets should have equal number of particles and
antiparticles, one should find more $p,\Lambda, K$ and $K^{\star}$
than ${\bar{p}}, \bar{\Lambda}, \bar{K}$ and
$\overline{K^{\star}}$ at mid-rapidity in heavy-ion collisions. It
was found that the flavor dependence and suppression factor would
be a good probe of the energy loss and other novel effects
\cite{wangflavor,vitev}. Fig.~\ref{Fig:kstarflavor} shows the
HIJING simulation of antiparticle to particle ratio as function of
$p_{T}$ at midrapidity ($|y|\!<\!0.5$) in Au+Au collisions at
$\sqrt{s_{_{NN}}}\!=\!200$ GeV. It can be seen that the ratios of
$\bar{K}/K$ and $\overline{K^{\star}}/K^{\star}$ have the same
sensitivity to the flavor dependence of particle production as
$\bar{p}/p$ and $\bar{\Lambda}/\Lambda$. In addition, the
$K^{\star+}$ measurement will allow the measurement of the isospin
effect by using $K^{\star+}/K^{\star}\!\simeq\!1.0$ and comparing
to the flavor dependence of $\overline{K^{\star}}/K^{\star}$.
Experimentally, the momentum resolution of lower $p_T$ particles
is better than higher $p_T$ due to the rigidity (curvature) in the
magnetic field (e.g. STAR). The reconstruction of two or three
lower momentum daughters to form a high $p_T$ particle is less
sensitive to the distortion when we want to study the flavor and
isospin effect. Experiments \cite{starlambda} have shown that the
ratios of $K^{+}/K^{-}$ and $\bar{\Lambda}/\Lambda$ at low $p_{T}$
do not depend on $p_{T}$. It is important to find the onset (on
$p_{T}$) of the strong flavor dependence as predicted.

In summary, studies of strongly interacting resonant states open new
approaches to the study of relativistic heavy-ion collisions.  We
found that the yields of resonances deviate from thermal fits at
chemical freeze-out. This is likely related to the evolution of the
system and continuous interactions after chemical freeze-out.  The
observation of $\rho^{0}$ and $K^{\star0}$ mass shifts in both $pp$
and Au+Au collisions are actively investigated theoretically and
experimentally. These results show clear dynamic effects at the
freeze-out. The transverse momentum spectra of resonances at
intermediate and high $p_T$ can help disentangle different effects.

The auther would like to thank the useful discussions with Drs. G. Brown, M. Bleicher, P. Fachini,
L. Gaudichet, T. Hallman, H. Huang, R. Longacre, J. Ma, R. Majka, C. Markert, R. Rapp, F. Retiere,
J. Sandweiss, E. Shuryak, A. Tai, T. Ullrich, N. Xu, H. Zhang. The references from LEP and NA27
were done together with P. Fachini. The UrQMD results were obtained from M. Bleicher
(during his visit to BNL) and P. Fachini.

\end{document}